\documentclass[conference]{IEEEtran}
\usepackage{todonotes}
\usepackage{graphicx}  % randy: allow .pdf for figures
\usepackage{float}  % randy: allow "[H]" (place here) for figs
\usepackage{cite}
\usepackage{listings}

\ifCLASSINFOpdf
\else
\fi
\usepackage{amsmath}
\usepackage[hyphens]{url}
\usepackage[normalem]{ulem}

\begin{document}
%
% paper title
% Titles are generally capitalized except for words such as a, an, and, as,
% at, but, by, for, in, nor, of, on, or, the, to and up, which are usually
% not capitalized unless they are the first or last word of the title.
% Linebreaks \\ can be used within to get better formatting as desired.
% Do not put math or special symbols in the title.

% \title{Bringing Containers to Scientific Workflows:\\The Pegasus Example}
\title{Custom Execution Environments with Containers\\in Pegasus-enabled Scientific Workflows}

\author{\IEEEauthorblockN{
    Karan Vahi\IEEEauthorrefmark{1}, 
    Mats Rynge\IEEEauthorrefmark{1},
    George Papadimitriou\IEEEauthorrefmark{1},
    Duncan A. Brown\IEEEauthorrefmark{2},
    Rajiv Mayani\IEEEauthorrefmark{1} \\
    Rafael Ferreira da Silva\IEEEauthorrefmark{1},
    Ewa Deelman\IEEEauthorrefmark{1},
    Anirban Mandal\IEEEauthorrefmark{3},
    Eric Lyons\IEEEauthorrefmark{4}, 
    Michael Zink\IEEEauthorrefmark{4}
    }
    
    \IEEEauthorblockA{
        \IEEEauthorrefmark{1}Information Sciences Institute, University of Southern California, Marina Del Rey, CA, USA \\
        \IEEEauthorrefmark{2}Department of Physics, Syracuse University, NY, USA \\
        \IEEEauthorrefmark{3}RENCI - University of North Carolina, NC, USA \\
        \IEEEauthorrefmark{4}Department of Electrical and Computer Engineering, University of Massachusetts at Amherst\\
        \{vahi,rynge,georgpap,mayani,rafsilva,deelman\}@isi.edu, \\
        dabrown@syr.edu,anirban@renci.org,elyons@engin.umass.edu,mzink@cas.umass.edu
    }
}

% use for special paper notices
%\IEEEspecialpapernotice{(Invited Paper)}

% make the title area
\maketitle

% As a general rule, do not put math, special symbols or citations
% in the abstract
\begin{abstract}
Science reproducibility is a cornerstone feature in scientific workflows. In most cases, this has been implemented as a way to exactly reproduce the computational steps taken to reach the final results. While these steps are often completely described, including the input parameters, datasets, and codes, the environment in which these steps are executed is only described at a higher level with endpoints and operating system name and versions. Though this may be sufficient for reproducibility in the short term, systems evolve and are replaced over time, breaking the underlying workflow reproducibility. A natural solution to this problem is containers, as they are well defined, have a lifetime independent of the underlying system, and can be user-controlled so that they can provide custom environments if needed. This paper highlights some unique challenges that may arise when using containers in distributed scientific workflows. Further, this paper explores how the Pegasus Workflow Management System implements container support to address such challenges.
\end{abstract}

% no keywords

% For peerreview papers, this IEEEtran command inserts a page break and
% creates the second title. It will be ignored for other modes.
\IEEEpeerreviewmaketitle

%% Sections
\section{Introduction and Motivation}

Container technologies have become ubiquitous in all areas of scientific computing~\cite{boettiger2015introduction, di2015impact}. While Docker~\cite{docker} has been a leader in areas such as microservices, containers did not become common in high throughput computing (HTC) or in high performance computing (HPC) environments until singularity~\cite{kurtzer2017singularity} was introduced. Scientific computing specialists now have a range of container tools available to them, in addition to Docker and Singularity. There are also HPC-specific solutions like Shifter~\cite{shifter}, open frameworks such as Podman~\cite{podman} (which can be used to develop, manage, and run application containers), and orchestration tools like Kubernetes~\cite{kubernetes} (which can be used for automated deployment, scaling, and management of containerized applications).

In the context of scientific workflows, container technologies are especially interesting for two reasons: (1)~they supply a way to foster reproducibility by providing a fully defined and reproducible environment; and (2)~they are able to provide a flexible, custom, user-controlled environment in a manner that the underlying centrally managed compute cluster cannot (due to the fact that administrators' have a main goal of providing a slow-moving, stable, multi-user environment). Science reproducibility in scientific workflows is often implemented as a way to reproduce the exact computational steps taken to reach the final results. While these steps are completely described, including the input parameters, datasets, and codes, the environment in which these steps are executed is only described at a higher level with endpoints and operating system name and versions. While this is enough to attain reproducibility in the short term, systems evolve and are replaced over time, breaking the workflow reproducibility. Containers are well defined and have whatever lifetime the owner desires, meaning that containers outlast whatever compute environment the workflow was first executed in. Similarly, containers also enable the workflow to be seamlessly transferred to completely different compute environments.

Applications are increasingly relying on a diverse set of underlying technologies and libraries to optimize the use of evolving computing hardware. As a result, requests for custom software environments are becoming more and more common. Users' software stack requirements may conflict with a system-provided stack or even with other users. These requirements are often impossible to satisfy for a stable, multi-user compute cluster environment. 
%A key reason is that some areas of computing are advancing at a very quick pace, and therefore the software stack requirements are commonly evolving faster than a stable multi-user compute cluster environment can provide. 
A typical example nowadays is TensorFlow~\cite{abadi2016tensorflow}, a popular machine learning toolkit. TensorFlow is Python-based and requires a set of very recent Python libraries. This environment is not impossible to satisfy under a for example, RHEL 7 based compute cluster. However, TensorFlow is a non-trivial stack to build and provide support for. Many HPC centers now solve this problem by providing Singularity images of such tools.

While containers increase reproducibility and enable custom environments, using containers in a distributed scientific workflow introduces some unique challenges. Most notable is the challenge of distributing the associated container images and making them available to the compute jobs. Pegasus~\cite{deelman-fgcs-2015} workflows regularly contain thousands or millions of jobs, simultaneously running across a set of different compute environments. To distribute the image at such a large scale and make the image available on the node where a job executes, special care has to be taken. Additionally, the container technologies are fragmented, so a one-size-fits-all approach may not be ideal when trying to support workflow execution in varied execution environments. 

With the above goals in mind we have added new capabilities to Pegasus WMS to support variety of Container technologies in different execution environments.

%More specifically, this work makes the following contributions:
%\begin{enumerate}
%    \item 
%\end{enumerate}

This paper is organized as follows. We start by describing the requirements and design considerations that we identified to support containers in Pegasus. Our approach to support containers is described in Section~\ref{sec:approach}. In Section~\ref{sec:experiments}, we present results from our experiments on a real world workflow and we quantify some overheads that result from using containers in workflows.  Section~\ref{sec:casestudy}  describes two application workflows that are now using containers for their production runs. In Section~\ref{sec:casestudy}, we also report our experiences with and lessons resulting from executing these workflows with containers in a production environment. Section\ref{sec:relatedwork} describes the related work and finally, Section~\ref{sec:conclusion} concludes with a brief summary of results and a discussion of potential future research.

% \begin{itemize}
%     \item The use of containers has become popular in the past decade
%     \item Motivation
%     \item Easiness of deployment
%     \begin{itemize}
%         \item No need to depend on sysadmins to perform the installations and configurations
%         \item Mitigates software dependencies issues
%     \end{itemize}
% \end{itemize}

\section{Requirements and Design Considerations}
\label{sec:requirements}

Pegasus allows scientists to describe their computational pipelines as a directed, acyclic graph of tasks. The Pegasus input format (called the DAX) is a high-level, portable description that is agnostic of the underlying computing environment and refers to data and user codes using logical identifiers. Pegasus consumes this description and generates an executable workflow that is tailored for users that target computing environments such as local desktops, campus clusters, computational grids, and cloud environments. During this process, Pegasus automatically identifies the necessary input data and adds data management tasks to the user workflow that are responsible for fetching the data required for the workflow, stages-out the generated outputs to a user specified location, and optionally removes data products that are no longer required as the workflow executes.  The separation between the high-level user description of the pipeline in the DAX and the actual executable workflow that is executed on the computing resources has enabled our user community to keep abreast with infrastructure improvements and migrate their pipelines from original HPC focused environments (such as local campus clusters) to more distributed computing environments (such as Open Science Grid and clouds).

Running an application or service using containers is a well known and straightforward process. However, when incorporating this process into scientific workflow systems presents a unique set of challenges. Thus, we identified some over-arching requirements for architecting support for containers in Pegasus. There requirements are listed below:

\begin{enumerate}
\item \emph{Support for different container technologies}: Early on it was clear to us that a one-size-fits-all approach would not suffice when picking a container technology for widespread use. For example, Docker, while popular in traditional corporate computing environments and in local captive computing resources, is not supported on most shared computing infrastructures as the Docker agents and jobs run as root. Singularity, however, is a preferred container technology that allows containerized jobs to run in user space in HPC environments. Some HPC centers such as National Energy Research Scientific Computing Center (NERSC) have introduced their own specific container technologies, like Shifter, that enable users to securely run Docker images on NERSC systems at scale. The goal of Pegasus is to allow users to optimize computing resources at their disposal and leverage technologies that are supported on those resources.
\item \emph{Work in Distributed Environments}: Irrespective of the container technology supported, it was important to ensure that Pegasus' support enabled users to utilize containerized jobs in distributed environments. In such environments, users often don't know \textit{a priori} which node or cluster their job might land on. The task of fetching and deploying a container that a job requires on a node and any associated setup required for the container (such as loading an image in the local container registry, etc.) should be handled by the workflow management system and not in users' scripts. 
\item \emph{Easy Configuration and Representation}: It should be easy for users to configure which container and type of container their jobs require, allowing different jobs in the same workflow to use different containers and technologies. The underlying representation to describe containers used for a workflow should be compact and prevent duplication.
\item \emph{Support for Public Registries}: Today, a lot of popular container images are available to users in public registries (such as Docker Hub and Singularity Hub). Our solution should support retrieval of images from these registries for use in a workflow and should also scale-up for large workflows whereby the registries are not accessed repeatedly for the same image. Private images can be loaded directly from an image file.
\end{enumerate}

\section{Approach}
\label{sec:approach}

%
%\begin{itemize}
%    \item Maybe name as Model and Design
%    \item Representation
%    \item Data Management of container
%    \item Symlinking
%    \item Container Execution Model
%    \item Inside user
%    \item Mount points
%\end{itemize}

We based our overall approach to incorporating support for containers by making containers first class citizens in our model and treating them as an input dependency for a job. Keeping in mind that we wanted our solution to scale-up for large workflows and access public registries without overloading these registries during a workflow run, we decided to represent the container dependency as a data dependency, and leverage Pegasus data management capabilities to manage distribution of containers required for a workflow. Support for containers in Pegasus was first introduced in Pegasus 4.8.0, which was released in September 2017 with support for Singularity and Docker Containers. The current version of Pegasus, version 4.9.1, also has support for Shifter.

\subsection{Container Execution Model}
\label{sec:container-exec-model}
Executing a job via a container on a remote node usually requires some setup and cleanup actions before and after the job has run. For example, before launching a job, the associated container image might need to be retrieved and loaded in the local container registry. After job completion, the container image might need to be unloaded/removed.
We decided to incorporate the container setup for a job into PegasusLite\cite{vahi2013rethinking}, a light-weight Pegasus remote execution engine which wraps the user task on the remote worker node when a job is scheduled to the node. PegasusLite is responsible for figuring out the appropriate job directory in which the job executes, staging-in datasets that a job requires, launching the job, staging-out data, and cleaning up the job directory. 

\begin{figure}[H]
	\centering
	\includegraphics[scale=0.4]{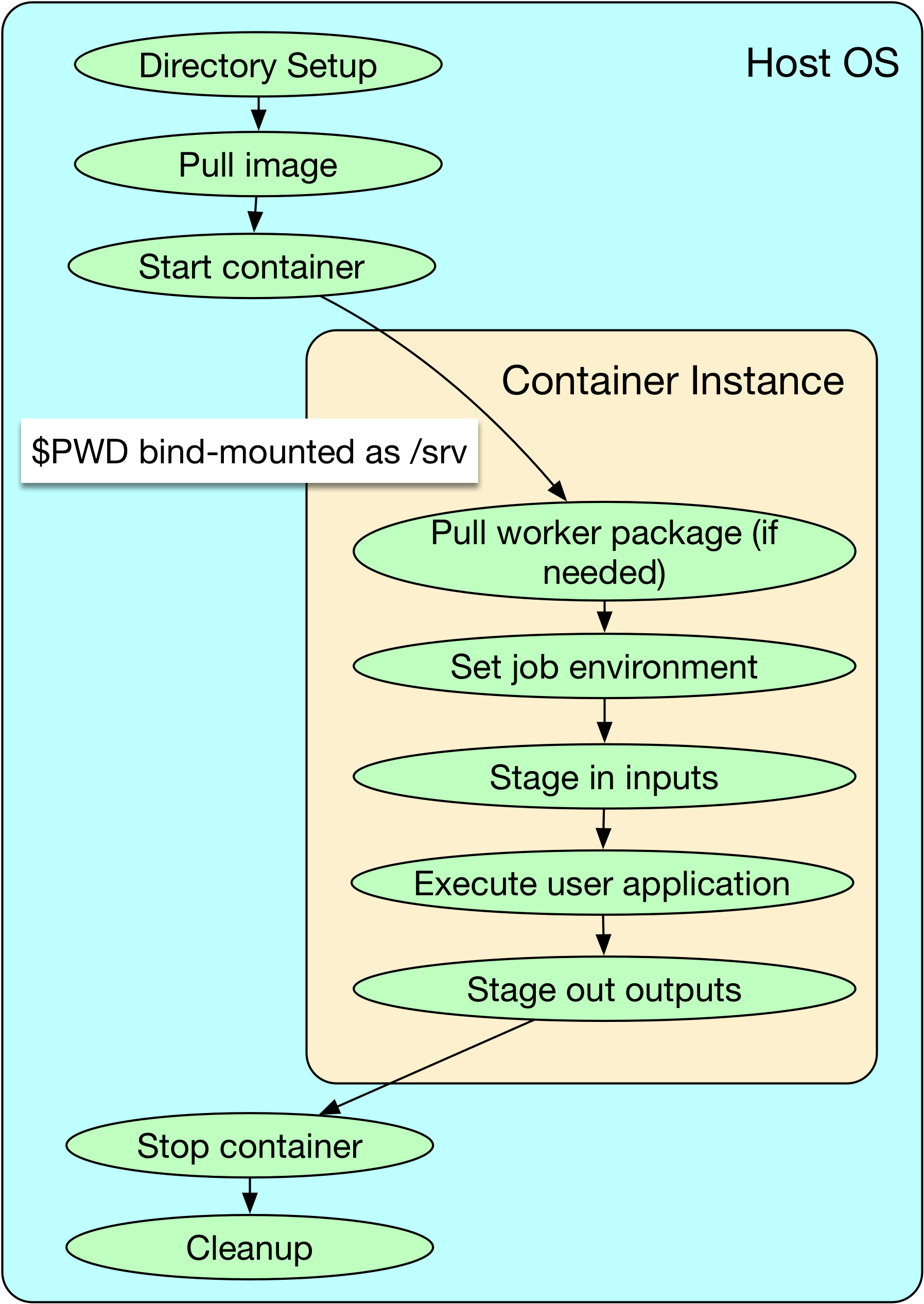}
	\caption{The PegasusLite steps taken on the bare-metal host and inside the container instance.}
    \label{fig:container_host}
\end{figure}

The updated PegasusLite flow for handling a containerized job when starting on a remote worker node is outlined below:

% \todo[inline]{I think it is important to point out where the steps takes place host vs container - maybe we need a figure here}

% \todo[inline]{Don't we stage out from inside the container instance? If so 4e and 5 are swapped} . Fixed now by Karan

\begin{enumerate}
    \item Sets up a directory in which it will run a user job.
    \item Pulls or links the container image to that directory.
    \item Mounts the job directory into the container as /scratch for Docker containers, while as /srv for Singularity containers.
    \item Container then runs a job specific script created by PegasusLite that does the following:
    \begin{enumerate}
        \item Figures the appropriate Pegasus worker to use in the container if not already installed.
        \item Sets up the job environment to use (including transfer and setup of any credentials transferred as part of PegasusLite).
        \item Pulls in all the relevant input data and  executables required by the job.
        \item Launches the user application using pegasus-kickstart.
        \item Ships the output data to the staging site.
    \end{enumerate}
    \item Optionally, shuts down the container (only applicable to Docker containers).
    \item Cleans up the directory on the worker node.
\end{enumerate}

The Pegasus planner configures PegasusLite for each job, with container specific directives using a pluggable interface. This interface allows us to  easily incorporate newer container technologies. Below, we describe the supported Container technologies and how PegasusLite is configured for each of them. For the purposes of this section, we assume the associated container image file is present on the node where the job is executed.  Section~\ref{sec:container-data-management} describes how the container image required for the job gets to the worker node.

\subsubsection{Docker}

PegasusLite first loads the container from the container image file into the local node registry. It then identifies the user (the user on the host OS) that the job is being launched as and creates the same user in the container if it does not exist. Running the job inside the container with the same UID/GID as the host OS ensures data access to the host working directory and guarantees any outputs created by the job are as the same user as on the host OS. The job directory as determined by PegasusLite gets mounted as /scratch into Docker containers.  The job is then setup and executed in the container as described in Step 4.
    
\subsubsection{Singularity}

Comparatively, Singularity setup is more straightforward as it is designed to be executed in user space and the image file can be invoked as any other Linux executable using singularity exec command. In case of Singularity, the job directory (as determined by PegasusLite) is mounted as /srv in the container, and then the job is setup and executed in the container as described in Step 4. Using /srv under Singularity comes from Singularity not using overlay-fs in some cases, and hence having to rely on existing mount points in the image when bind mounting. 

\subsubsection{Shifter}

Shifter is a technology developed  by NERSC and optimized for executing on HPC machines. Shifter containers are different from Docker and Singularity because Shifter containers cannot be exported to a container image file that can reside on a filesystem. Additionally, the containers are expected to be available locally on the compute sites in the local Shifter registry. In case of Shifter, the job directory (as determined by PegasusLite) is mounted as /scratch in the container and the job is then setup and executed in the container as described in Step 4.

\subsection{Data Management for Containers}
\label{sec:container-data-management}
Pegasus treats containers as an input data dependency for a job that needs to be staged to a compute node if it is not already present there. If a container is described in the Transformation Catalog as residing in a container registry (such as Docker Hub or Singularity Hub), we first export the image as a container image file as part of the executable workflow. To achieve this, we modified our data transfer tool \textit{pegasus-transfer} to pull the images from Docker or Singularity Hub and export them as container image files. Treating containers as data allows us to:
\begin{itemize}
    \item Execute wokflows in distributed environments, where the container, along with job input data, gets deployed at runtime on a remote compute node when a job starts.
    \item Optimize transfer of containers for large workflows in a manner similar to how Pegasus does for datasets\cite{vahi2013rethinking}. This is of particular importance when workflows refer to containers in public container registries. For a workflow, Pegasus will retrieve a particular container image from a public registry once per data staging site irrespective of the number of jobs in the workflow. This is significant as the access pattern for container images from a workflow execution can appear as a Denial of Service attack for the operators of the registries.
    \item Export the image to a file format from a Hub, which also allows us to stage-in the image to the compute nodes via a staging server in the instance that the actual compute nodes may not have direct access to the public internet.
    \item Symlink against a container image file if a shared filesystem is available on the compute nodes of a target execution resource. In this case, the data staging node added by Pegasus in the executable workflow will place the container image on a directory on the shared filesystem and all the jobs will then symlink the container into their job directories. This is particularly useful when jobs refer to large container images. This also minimizes saturation of network links in a compute site. In Section~\ref{sec:experiments}, we highlight the benefit of this optimization.
\end{itemize}

By default, Pegasus only mounts the job directory determined by PegasusLite into the application container. However, in the Transformation Catalog, users can specify additional directories that need to be mounted into the container and made available to the job. This allows jobs in the workflow to symlink against pre-existing input data sets that may be available on the node. Many computing sites now distribute datasets using CVMFS~\cite{osg-cvmfs} (e.g., the Gravitational-wave Open Science Center~\cite{Vallisneri:2014vxa}), which Pegasus can mount into the container automatically when a user job starts.

Overall, treating containers as an explicit data dependency for jobs has given us the flexibility to leverage containers in a variety of execution environments while simultaneously optimizing the access pattern based on the individual compute environments.

\subsection{Representation}
\label{sec:container-representation}
Pegasus users describe their applications as transformations in a Transformation Catalog. The Transformation Catalog maps logical transformations to physical executables on a particular system. It also provides additional  information about the transformation such as what system they are compiled for, what profiles or environment variables need to be set when the transformation is invoked, and so on. Users have an option of marking the executables as installed on a particular system or as stageable, in which case Pegasus will transfer the executable along with the job.  We updated the Transformation Catalog to allow users to refer to a container that is required for execution. A sample representation is illustrated below

\lstset{
    basicstyle=\ttfamily\scriptsize,
}
\begin{lstlisting}
- transformations:
  - namespace: "example"
    name: "keg"
    version: 1.0
    site:
    - name: "isi"
      arch: "x86"
      os: "linux"
      container: "centos-pegasus"
      pfn: "/shared/pegasus/bin/pegasus-keg"
      
      # INSTALLED means pfn refers to path in the container.
      # STAGEABLE means the executable can be staged into
      # the container
      type: "INSTALLED"
      
- cont:
  - name: "centos-pegasus"
    
    # URL to image in a docker|singularity hub|shitfer 
    # repo url OR URL to an existing docker image 
    # exported as a tar file or singularity image
    image: "docker:///rynge/montage:latest"
  
    # can be either docker or singularity or shifter
    type: "docker"
    
    # mount information to mount host directories into 
    # container  format src-dir:dest-dir[:options]
    mount:
    - "/Volumes/Work/lfs1:/shared-data/:ro"
   
    # environment to be set when the job is run in the 
    # container only env profiles are supported
    profile:
    - env:
        JAVA_HOME: "/bin/java.1.6"
\end{lstlisting}{}

The container itself is defined using a separate cont entry. Multiple transformations can refer to the same container. Users can choose whether to use either the same container for all their jobs in the workflow or different containers for different types of jobs. We also support the notion of allowing users to stage their executables into a container at runtime. This is useful when a user maybe using a standard base container image from a public repository but wants to user their own executables.

%\todo [inline]{do we need to elaborate about executable staging, as in statically linked or dynamicallly linked as long as libraries exist in the container}

We briefly describe the attributes supported for describing a container

\begin{itemize}
    \item cont - A container identifier.
    \item image - URL to image in a docker|singularity hub| shifter repo URL or URL to an existing docker image exported as a tar file or singularity image. An example of a docker hub URL is docker:///rynge/montage:latest and an example of a singularity is shub://singularity-hub.org/pegasus-isi/fedora-montage. Shifter images can only be referred to by a shifter URL scheme that indicates that the image is available in the local shifter repository on the compute site. An example of this is shifter:///papajim/namd\_image:latest.
    \item mount - mount information to mount host directories into container of format src-dir:dest-dir[:options]. This is used to mount directories from the shared filesystem on a compute site in the container, for symlinking against pre-existing inputs a job may require.
    \item profiles - One or many profiles can be attached to a transformation for all sites or to a transformation on a particular site. For containers, only env profiles are supported.
\end{itemize}

\section{Experiments}
\label{sec:experiments}

In order to obtain a better understanding of the performance overheads when adding container transfers and instance management, a set of experiments were conducted. The Chameleon testbed\cite{chameleon-cloud} was used as the testbed, which consisted of one workflow submit node, one NFS server node, and four worker nodes. All of the nodes were bare metal nodes with 24 physical cores, 128GB of RAM, and 10Gbps network connection. However, because we wanted to simulate lower network speeds on the submit node, we capped its network link to 1Gbps. Additionally, the submit node didn't have access to the storage server via NFS. The shared filesystem was only shared across the worker nodes. Our software stack on the submit node apart from HTCondor and Pegasus, also included an {\it http} server, in order to allow the workers to fetch input data to the jobs. 

\begin{figure}
	\centering
	%\scalebox{0.9}{\input{plots/storage_skylake/makespan_all_run_types.tex}}
	\includegraphics[width=0.9\linewidth]{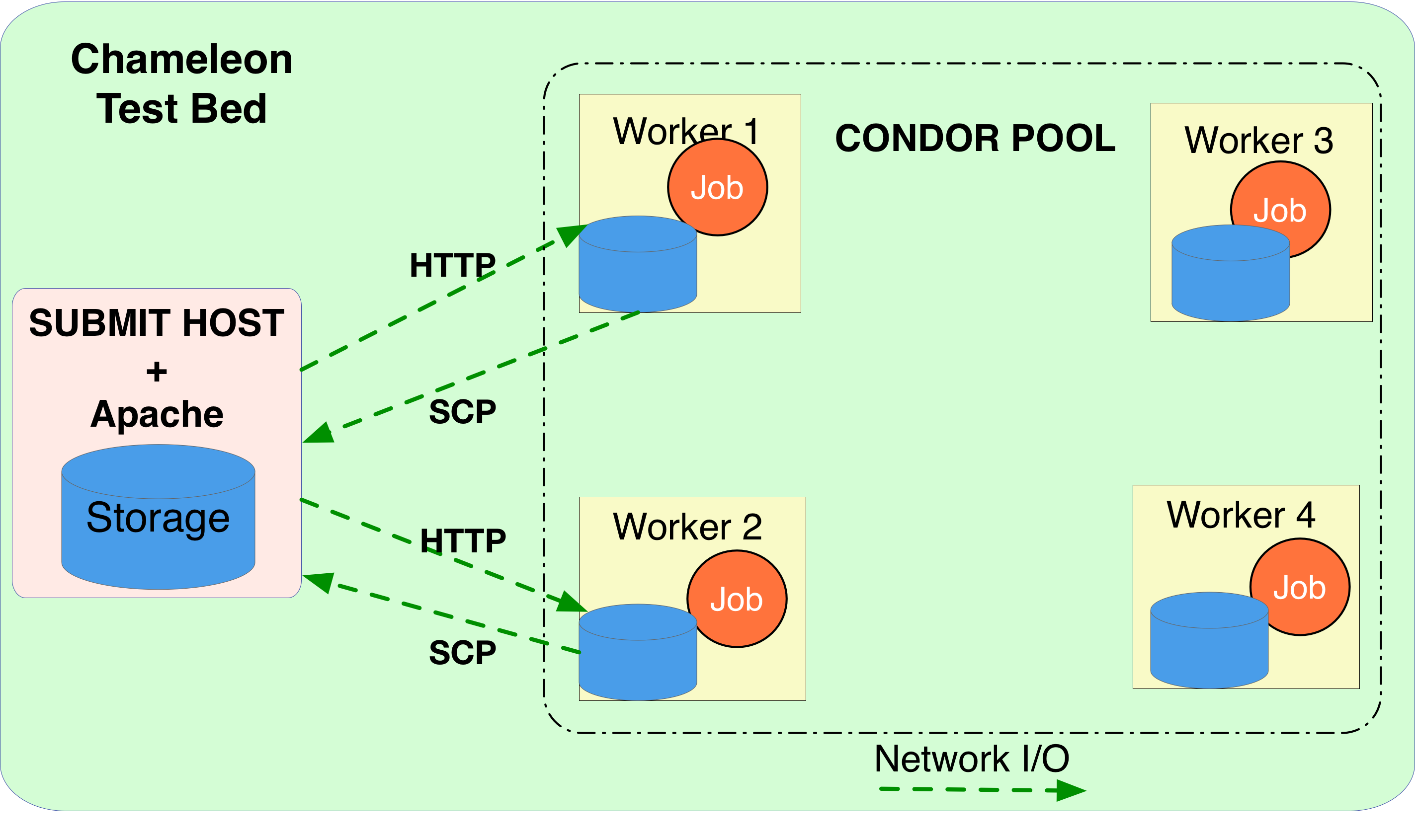}
	\caption{Non Shared Fileystem Setup\vspace*{-0.1in}}
	\label{fig:vanilla_setup}
\end{figure}

\begin{figure}
	\centering
	%\scalebox{0.9}{\input{plots/storage_skylake/makespan_all_run_types.tex}}
	\includegraphics[width=0.9\linewidth]{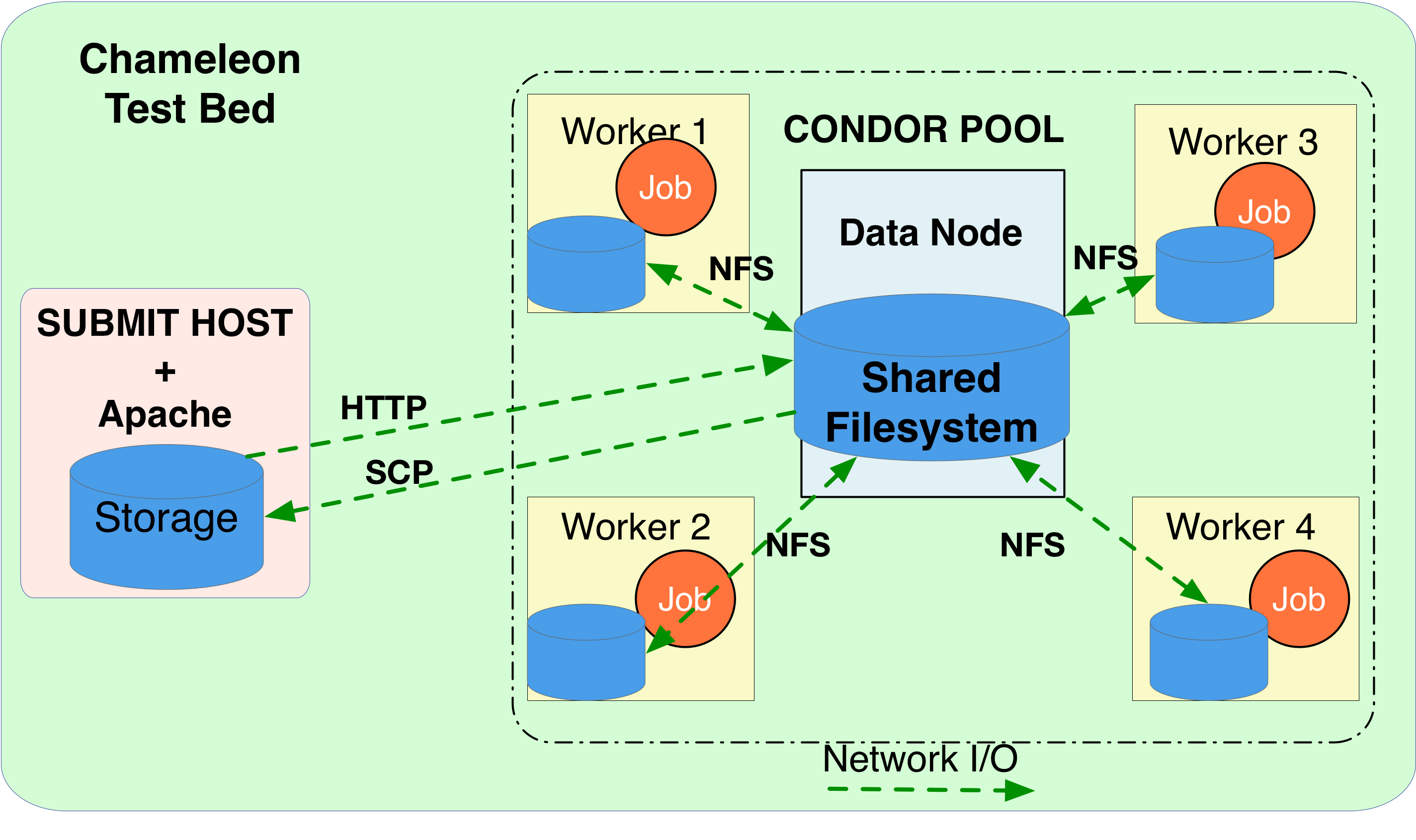}
	\caption{Shared Filesystem Setup\vspace*{-0.1in}}
	\label{fig:shared_setup}
\end{figure}

%\todo[inline]{George update description of testbed}
As a test workflow, we used the CASA application workflow described in Section\ref{sec:casa}, which is normally configured to use Docker containers. The workflow was configured to use {\it http} for staging-in data to the jobs and {\it scp} to store data back to the data staging site. The CASA workflow we used in our experiment had 63 compute tasks in the abstract workflow and 73 jobs in the executable workflow (when there is no task clustering). The 10 additional tasks are the data transfer and auxiliary tasks added by Pegasus during the planning of the workflow. Due to the real-time nature of this workflow, each compute task is designed to complete execution within a few seconds as it processes incoming data every minute.

We devised three experiments. All experiments executed the workflow using PegasusLite\cite{vahi2013rethinking} mode. In this mode, each job is wrapped in a lightweight PegasusLite instance that determines the directory in which the user task will be run, pulls in the input data for the job from data staging site, launches the task, and stages out the generated outputs back to the data staging site before cleaning up. The first experiment which served as a base, involved executing the CASA workflow without any containers, with input data (application data) staging as shown in Figure \ref{fig:vanilla_setup} occurring via {\it http} from the submit host. The second experiment involved executing workflows with containers (Docker and Singularity both), with input data (container and application data) staging as shown in Figure \ref{fig:vanilla_setup} occurring via {\it http} from the submit host, and the stage-out of data from workers to submit host via {\it scp}. In the third experiment, we staged the input data to the NFS (as shown in Figure \ref{fig:shared_setup}) and then had the compute jobs symlink against the input data on the NFS instead of storing a true copy to the local filesystem. For all the experiments we ran workflows 10 times in each configuration, and present average results over these 10 runs where applicable. Additionally, while running these experiments, in order to collect network and block device statistics, on all nodes we were recording system activity every second via {\it sar}\cite{sar}. The goals of these experiments were the following:

\begin{itemize}
    \item Demonstrate the increase in walltime due to the staging of application containers and how job clustering can help mitigate the overhead.
    \item Show that the staging of application containers for a workflow for each task can saturate both the network links and disk IO.
\end{itemize}

\begin{figure}
	\centering
	%\scalebox{0.9}{\input{plots/storage_skylake/makespan_all_run_types.tex}}
	\includegraphics[width=0.9\linewidth]{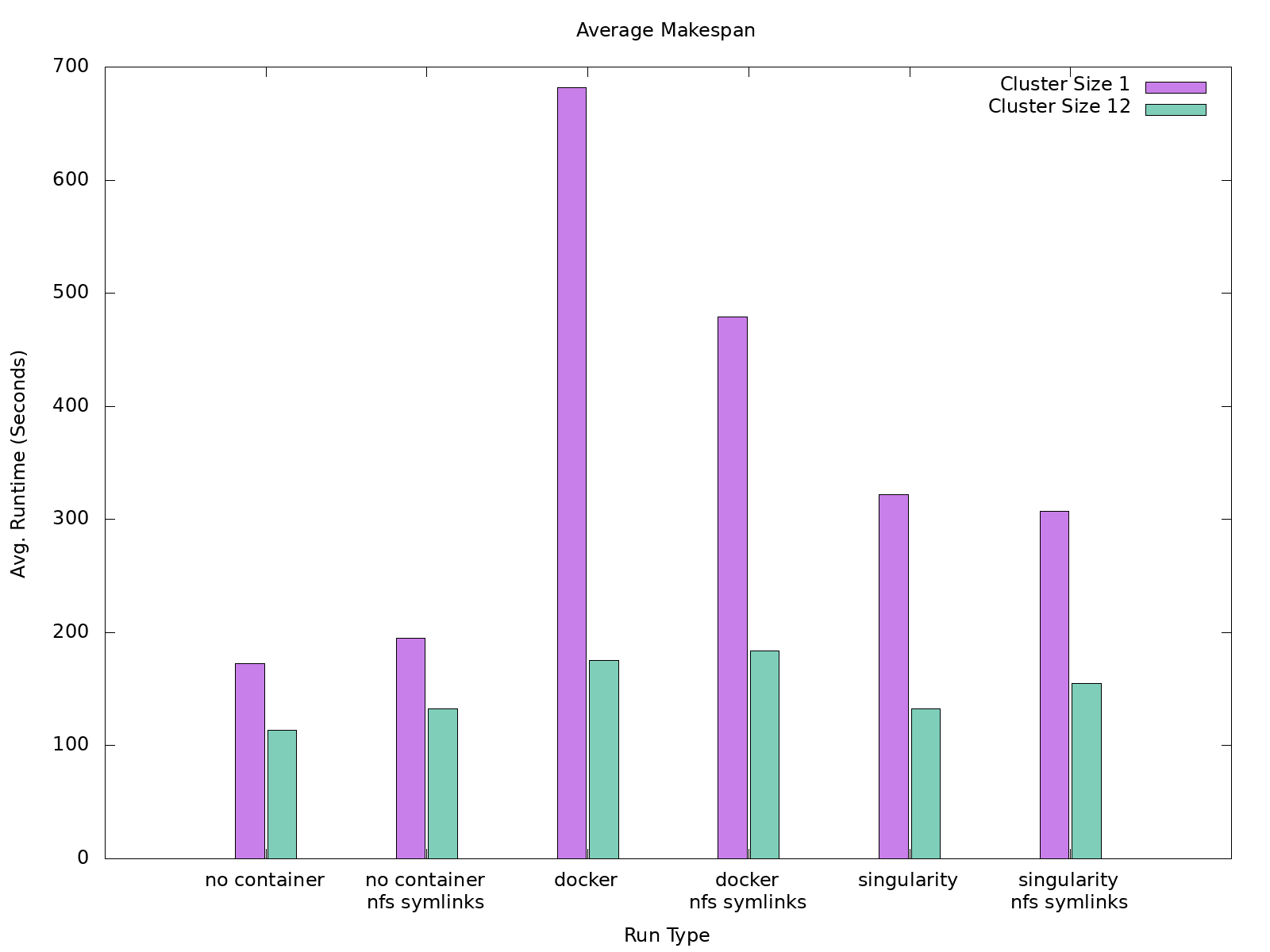}
	\caption{Average workflow makespan per execution environment setup\vspace*{-0.1in}}
	\label{fig:nowcast_makespan}
\end{figure}

The first graph in Figure \ref{fig:nowcast_makespan} shows an increase in end to end workflow walltime from 172.2 seconds to 681.7 and 321.6 when Docker containers and Singularity containers are used respectively and when there is no job clustering (Cluster size 1 - purple bar and lines in the plots). We also noticed that the Docker workflow walltime is much longer than Singularity. This is primarily because of the difference in the size of the Docker and Singularity image files (488MB and 153MB respectively) even though the underlying recipe files are functionally equivalent. Clustering the tasks together (so that one job executes 12 tasks; green bar and Lines in the plots) helps reduce the workflow walltime as job clustering leads the image being transferred only once per 12 tasks, and to fewer jobs in the final executable workflow; thereby reducing the number of times the container image is transferred from the submit host to the worker nodes.

\begin{figure}
	\centering
	%\scalebox{0.9}{\input{plots/storage_skylake/makespan_all_run_types.tex}}
	\includegraphics[width=0.9\linewidth]{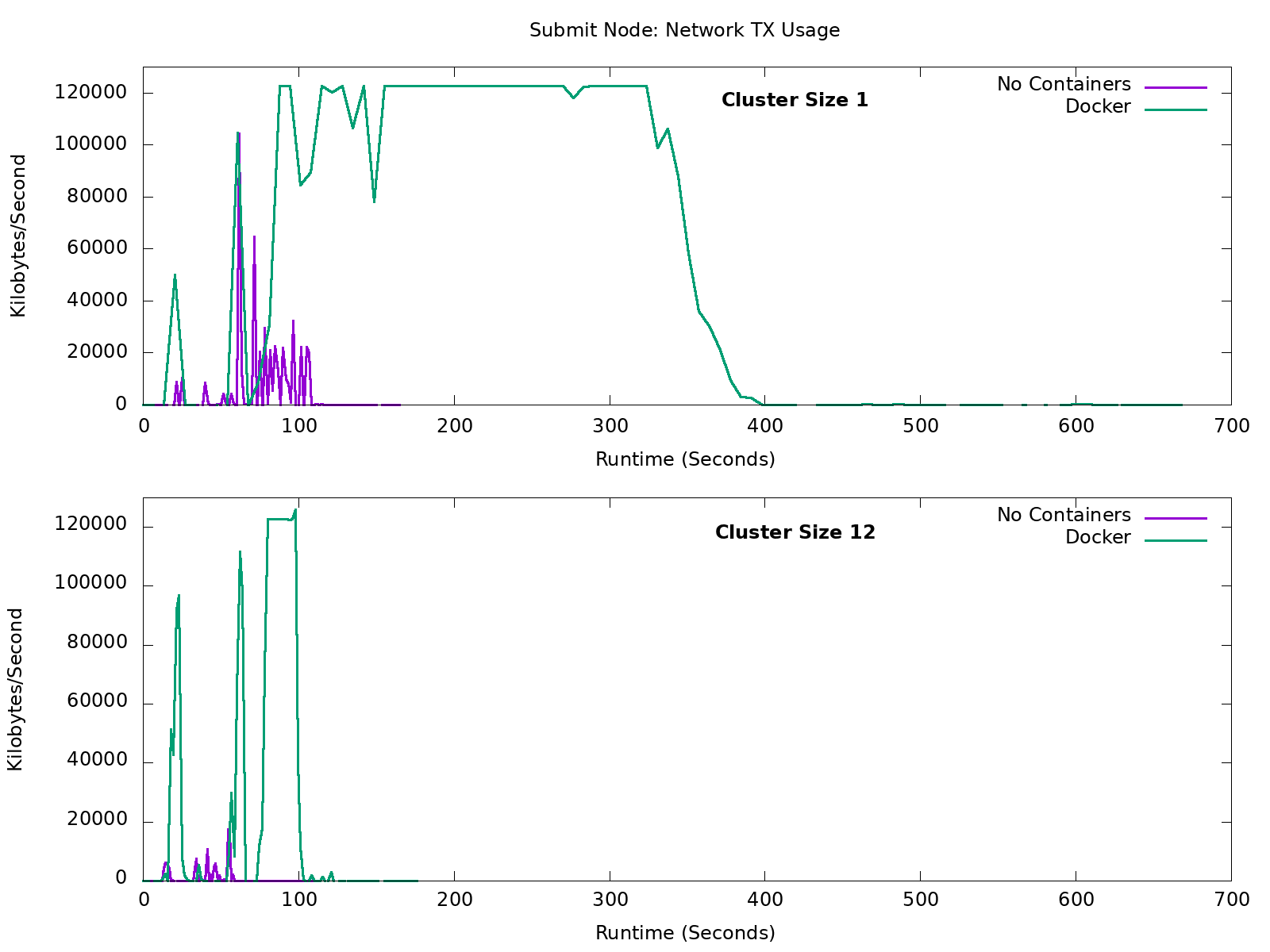}
	\caption{Egress network traffic on the submit node, without the use of containers and using Docker. NFS is \textbf{not} used.\vspace*{-0.1in}}
	\label{fig:nowcast_submit_network}
\end{figure}

Figure \ref{fig:nowcast_submit_network} shows the network link usage on the submit host, which is also the data staging site for the Docker case in shared filesystem setup as shown in Figure \ref{fig:shared_setup} (with and without job clustering). The graph shows sustained period of network saturation on the link because of the associated data transfers of the containers per job. With job clustering, the network saturation is not sustained.

\begin{figure}
	\centering
	%\scalebox{0.9}{\input{plots/storage_skylake/makespan_all_run_types.tex}}
	\includegraphics[width=0.9\linewidth]{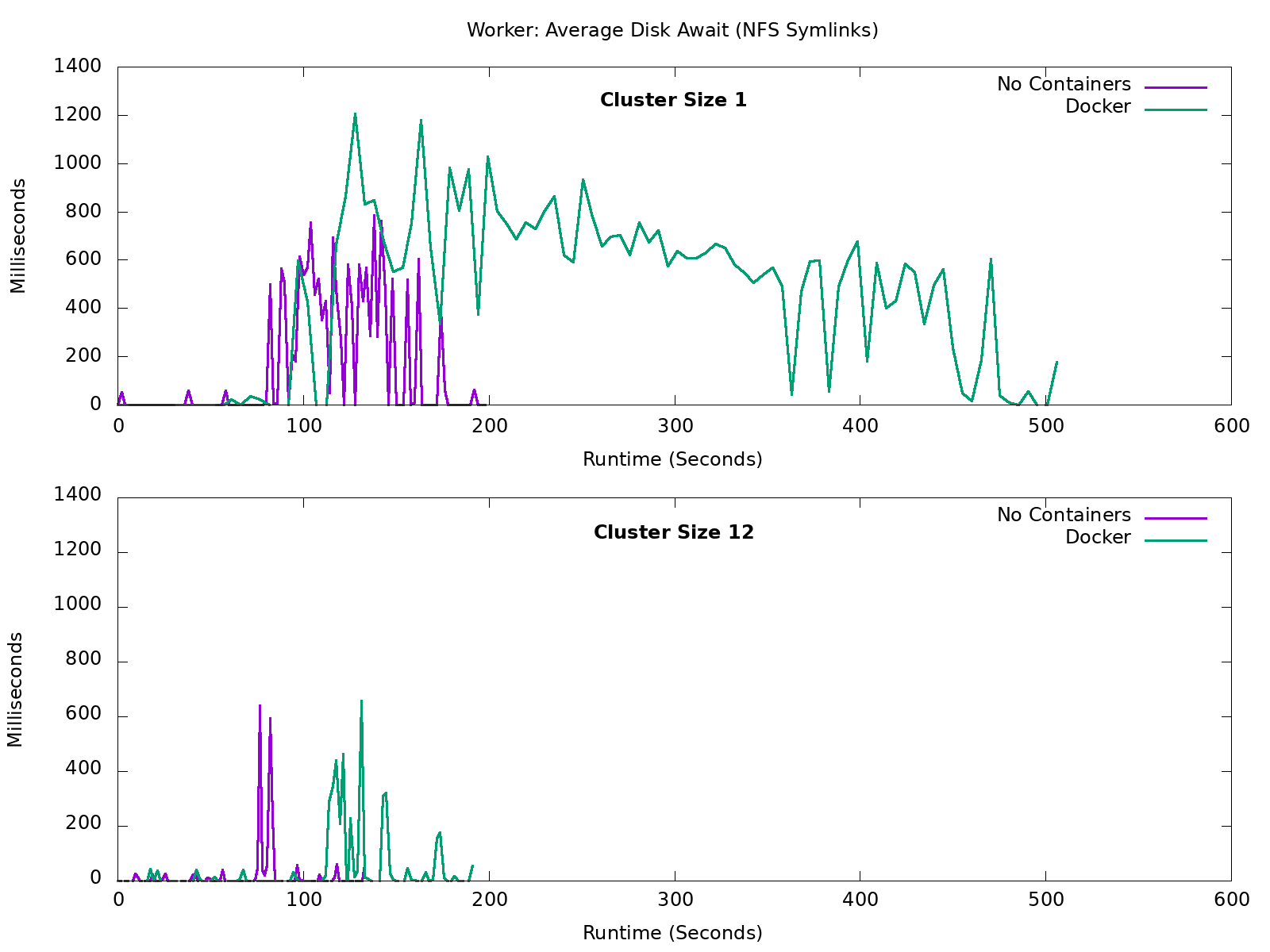}
	\caption{Average service time of I/O requests on worker 4 using Docker containers with NFS symlinking.\vspace*{-0.1in}}
	\label{fig:nowcast_worker_disk}
\end{figure}

\begin{figure}
	\centering
	%\scalebox{0.9}{\input{plots/storage_skylake/makespan_all_run_types.tex}}
	\includegraphics[width=0.9\linewidth]{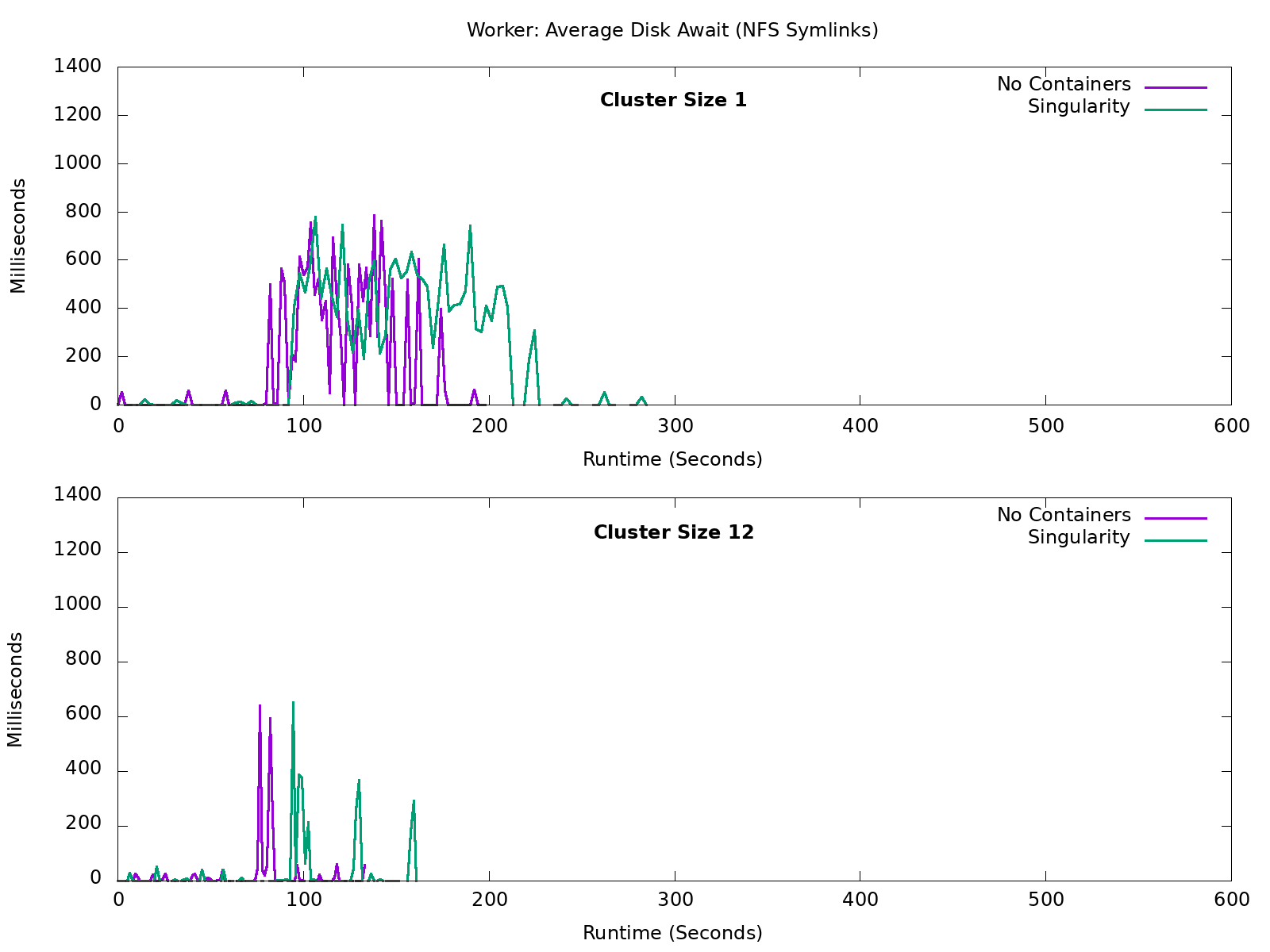}
	\caption{Average service time of I/O requests on worker 4 using Singularity containers with NFS symlinking.\vspace*{-0.1in}}
	\label{fig:nowcast_worker_disk_singularity}
\end{figure}

Figure\ref{fig:nowcast_worker_disk} shows average service time of I/O requests on one of the workers over a workflow run for runs with and without Docker containers; and with and without job clustering. In case of no containers cases, the effect on the average service time is negligible. Introducing containers leads to an increase in average disk wait times as container, even though we are symlinking against the Docker file on the NFS instead of doing a true copy to the local disk. The reason for this is that the Docker image image file still needs to be untarred internally on local disk by Docker before being loaded in the local node registry. A proposed optimization for smarter Docker loading in the PegasusLite script as explained in Section \ref{sec:conclusion}, which would lower the loading overhead when multiple jobs land on the same compute nodes. The same average service time plot for Singularity shown in Figure \ref{fig:nowcast_worker_disk_singularity} does not show a similar increase. This can be attributed both to Singularity images are read directly and are comparatively much smaller. 

\section{Case Studies}
\label{sec:casestudy}

Since container support was first added to Pegasus, our user base has been migrating their existing Pegasus pipelines to use containers. In this section, we describe two such applications, one using Singularity and the other using Docker.

% Subsection
\subsection{PyCBC}
\label{sec:pycbc}

PyCBC is a python-based software package used to explore astrophysical sources of gravitational waves~\cite{pycbc-github}. It contains algorithms that can detect coalescing compact binaries~\cite{Allen:2005fk,Allen:2004gu,Usman:2015kfa,Nitz:2017svb,Nitz:2018rgo} and measure the astrophysical parameters of detected sources~\cite{Biwer:2018osg}. PyCBC was used in the discovery of gravitational waves from binary black holes~\cite{Abbott:2016blz} and binary neutron stars~\cite{TheLIGOScientific:2017qsa}, in the ongoing analysis of gravitational-wave data by the LIGO and Virgo Scientific Collaborations~\cite{LIGOScientific:2018mvr}, and in the analysis of these data by independent groups~\cite{Nitz:2018imz}. 

PyCBC executables are Python scripts that call functions from PyCBC-provided, system, and third-party Python libraries, as well as compiled code from shared-object libraries. Unlike previous generations of gravitational-wave search executables that were provided as statically-linked C code with  minimal dependencies~\cite{Brown:2005zs}, PyCBC programs have a complex run-time environment that must be carried with each executable. A standard PyCBC installation requires that the install directory is available at runtime so that Python can find the package's libraries. It also requires that the build and runtime environments are compatible (e.g., compatible versions of glibc, gcc, and Python). Some PyCBC executables (e.g., \texttt{pycbc\_inspiral}) require runtime compilation of code using SciPy weave \cite{scipy}, so the execution environment must have a full installation of gcc and the Python development libraries.

On clusters where the user community has control over the Python interpreter and software installation, the complexities of the run-time environment have been managed by using a standard software installation and the  Python virtual environment tool \texttt{virtualenv} to create known environments into which all the software is installed. However, when running on OSG or XSEDE clusters for example, these requirements may not be satisfied. Although CVMFS can provide access to the PyCBC libraries at runtime on the OSG, many OSG execute machines do not have the required environment to weave-compile code at runtime. The initial solution for these environments was to build bundled executables using PyInstaller~\cite{pyinstaller}. These bundles contain all of the Python and user-space C libraries required, as well as a Python interpreter to run the code. This bundle must also contain pre-compiled objects for all of the weave code that is needed at runtime. A build script run under Travis CI as part of PyCBC's build and test system complied the necessary code into the bundle.

The use of PyInstaller allowed the construction of more self-contained executables that can be deployed on the OSG. However, this approach had several complexities: PyInstaller bundles are not completely static and still require a dynamically linked version of glibc at runtime. Since Linux systems are backwards but not forwards compatible, the bundle needed to be built on the lowest-common denominator operating system for the execution platform (e.g. RHEL6 for OSG). The bundle building script itself was quite complicated and the insertion of weave-compiled objects into the bundle required running the executable, extracting the complied object code from the weave cache, and inserting it into the bundle as part of the  build script. If a user added new code that was not exercised as part  of  this build, then the compiled objects would be missing from the bundled executable and attempts to use that feature would fail at run time.

Containerization has allowed us to mitigate these problems and create a more robust run-time environment. Since PyCBC's Travis CI build script was already configured to build a Docker container, this container is be converted into a Singulariy image using the OSG-provided  \texttt{cvmfs-singularity-sync} tool~\cite{osg-cvmfs-sync}. This tool pulls the latest PyCBC container (as well as released versions) from Docker Hub, converts the Docker container to a Singularity image, and copies the resulting image to the OSG's CVMFS origin server. Since the PyCBC maintainers have  complete control over the PyCBC Docker container environment, any necessary run-time software (including software needed for runtime compilation of code) can be installed into the container.

The implementation of containers in Pegasus meant that PyCBC's workflow generation script did not need to be modified as deployment of the container is managed by the Pegasus WMS. If the user wants Pegasus to run a PyCBC program inside a container, then Pegasus profiles are set to specify the container type, the path to the container image, and directives to mount CVMFS inside the container for access to data needed at runtime. The PyCBC program that invokes the Pegasus planner can set Pegasus profiles based on its command-line arguments, so no code changes were needed in the PyCBC code itself. During testing, we discovered that two changes were needed in Pegasus for optimal use of containers in PyCBC: better handling for containers distributed via CVMFS and better handling of data staging for containerized jobs. We describe these changes in Sec.~\ref{sec:lessons} below. A containerized PyCBC workflow was used to run the analysis for the First Open Gravitational-Wave Catalog~\cite{Nitz:2018imz} on the OSG.

% Subsection
\subsection{CASA}
\label{sec:casa}

CASA is an NSF Engineering Research Center with a focus on low atmosphere sensing, particularly with the use of networks of Doppler weather radars for the purpose of improving severe weather warning systems, emergency response, and situational awareness.  Radar systems produce voluminous data, requiring substantial computing and networking infrastructure to process the data in a timely manner.  Radar data is fused together to create derived products on which chains of post-processing occur, including image creation, GIS style analysis, and notification mechanisms for decision support.  One of these chains, Nowcasting, was selected as a representative example of a domain user workflow that was well tailored for this approach to processing.  Nowcasting is a form 
of short range forecasting that primarily uses observed radar data over time to produce a non-linear advection model that estimates future positions of radar echoes.  Every minute, the Nowcasting algorithm produces 31 grids, representing the projected radar reflectivity every minute from 0-30 minutes.  As part of the workflow, CASA creates images of each of these grids and extracts contours representing projected areas of meteorological impact based on predefined thresholds.  Contours are represented as GIS style polygons, which are effective ways of describing geographic areas of weather risk to end users. However, contouring into well-ordered non-convex polygons is a function of the weather contained and the grid size, which is a CPU intensive process approaching  O(${n}^2$) complexity .  Because timeliness is a key concern for effective end user response, innovative approaches to processing, including the use of the academic cloud, are necessary to keep up with the one minute update rate of nowcasting.

% Subsection
\subsection{Experiences/Lessons Learnt}
\label{sec:lessons}
Based on our users' feedback from using containers in their scientific workflows, we have introduced optimizations and made changes to our approach. Here we describe a few of these changes. 

\subsubsection{Direct Access to Singularity Images via CVMFS}

The PyCBC pipeline usually runs on a HTCondor-based computing infrastructure (e.g., on the Open Science Grid, on clusters at Syracuse University and AEI-Hannover, or on the LIGO Data Grid). On this infrastructure, the singularity image files are  distributed using CVMFS, a scalable, reliable, and low-maintenance software distribution service that is available on all the nodes. In the most general case, Pegasus opts to pull a container image once to the data staging site and then lets the jobs pull the image to the compute nodes filesystem along with the other input data. However, this approach did not take advantage of the out-of-band caching and distribution of the Singularity images provided by CVMFS. In order for PyCBC workflows to directly use Singularity images stored in CVMFS, we updated Pegasus to allow for bypassing of container images files to the staging site and to enable symlinking to pre-existing images on the compute sites. These improvements allowed PyCBC workflows to use containers without introducing any new data transfer overhead associated with container data-staging. 

\subsubsection{Moved transfer data staging into the container rather than the host OS}

In Pegasus 4.8.x and 4.9.0, the PegasusLite script (the lightweight job wrapper that is used to launch a job on the compute node) executing on the host OS was responsible for pulling in the job inputs from the staging site into the directory where the job runs, and then mounted that directory into the application container at runtime. The user application then was launched in the container and executed. The generated outputs were then staged back to the staging site by PegasusLite. This approach had the advantage of doing all the data transfers outside of the container relying on the tools provided by the computing infrastructure provider. However, this left the user with no control over using their own preferred choice of data transfer protocols when staging the data products to the worker nodes. With Pegasus 4.9.1, we moved data staging to occur within the application container so that users can install and leverage additional data staging tools specific to their infrastructure. We ran into this issue while executing PyCBC workflows on local cluster at Syracuse, where preferred grid transfer tools such as \textit{gfal-copy} are not pre-installed. 

\subsubsection{Loading multiple Docker image tar files}

Currently, Pegasus doesn't have any optimization in place to avoid loading an image that already exists in the local Docker images, or an image that is already getting loaded by a another PegasusLite script. This can result in multiple ``docker load -i'' commands to be dispatched within a short period of time, severely impacting the performance of the local disk. In Figure \ref{fig:nowcast_worker_disk}, we presented the effect of loading multiple Docker images during a CASA workflow run where we observed the average wait time for an I/O service request times to be as high as 1.2 seconds. As a result, a very powerful node (24 physical cores with 128GB RAM) becomes almost unresponsive. In a scenario that the workers are shared among projects, this can also affect others if their jobs get scheduled to the affected nodes. Preliminary results we have from machines using SSDs present a much more subtle effect, however we are planning to address this issue on the upcoming Pegasus release.
\section{Related Work}
\label{sec:relatedwork}

Over the past decade, container systems such as Docker and Singularity have emerged to facilitate the sharing and migration of software by defining immutable, reusable execution environments. Their rapidly adoption by the scientific community have already supported a number of scientific progresses~\cite{di2015impact, boettiger2015introduction}. Several systems have been leveraged to improve the way containers are stored, shared, and indexed. For instance, the OSG has leveraged the CVMFS filesystem for storing and sharing Singularity images for their users~\cite{osg-cvmfs}. This solutions provides a consistent and efficient environment across all OSG computing sites. On the deployment management aspect, Kubernetes~\cite{kubernetes}, an open source cluster manager for Docker containers, decouples application containers from the details of the system they run. Both technologies have significantly improved the way containers are built, stored, and delivered, and they have been leveraged by several workflow management systems as discussed below. Kubernetes has been widely used by EGI for managing containerized workloads and services. Through EGI Cloud Container Compute~\cite{egi-cc} users can start a cluster of virtual machines and create a Kubernetes cluster to run Docker containers, which can be spawned to their high throughput computing infrastructure.

In the context of scientific workflows, new systems have been designed for specifically running workflows on cloud environments with extended support for Linux containers~\cite{gerlach2014skyport, gerlach2015container, airflow, kotliar2018cwl, novella2018container, di2017nextflow}, while well-established workflow systems have evolved to provide seamless support for running containerized applications~\cite{zheng2015integrating, zheng2017deploying}. Skyport~\cite{gerlach2014skyport, gerlach2015container} utilizes Docker containers to solve software deployment issues via software isolation. It targets automated deployment of Docker containers on cloud environments. Once built, container images are stored in the Shock data management system~\cite{gerlach2014skyport}, in which data is represented as an object with a unique identifier (the Shock node ID) and metadata describing computational and scientific provenance information. At runtime, input files and Docker images are automatically deployed into cloud instances to perform computation. In Pegasus, we can emulate such behavior by running workflows on OSG where application container images are stored in the CVMFS filesystem. Additionally, we provide mechanisms to properly configure the execution environment within the container as defined by the user in the workflow description (as described in Section~\ref{sec:approach}). 

Airflow~\cite{airflow} is a workflow system for running DAG-based workflows on cloud platforms. Recently, they have enabled support for Kubernetes to orchestrate the execution of Docker-enabled application containers in commercial clouds. Similarly, CWL-Airflow~\cite{kotliar2018cwl} provides a lightweight abstraction layer for running workflows described with the Common Workflow Language (CWL)~\cite{amstutz2016common} interface. Pachyderm~\cite{novella2018container} is data-driven pipeline execution system natively built for running workflows on Docker and Kubernetes. In Pachyderm, workflows are organized into Git repositories and describe operations to be performed in data files stored in such repositories, which fosters reproducibility through version control. Nextflow~\cite{di2017nextflow} is a workflow management system that uses Docker and Singularity for multi-scale handling of containerized computation. More recently, they have also enabled support for Kubernetes in a similar manner as for Pachyderm.
% \todo[inline]{special attention should be paid to wf systems which are container native (e.g. Pachyderm)}
Most of these systems target the orchestration of application containers on cloud platforms. In contrast, our approach is platform agnostic, i.e. application containers can be deployed in a range of computing platforms including standard laptops, campus clusters, and HPC and HTC systems. To the best of our knowledge, Pegasus is the first workflow system providing such versatility while preserving the fundamental concept of scientific workflow portability, i.e. framework and platform agnostic descriptions of workflows.

% \todo[inline]{EGI seems to do a lot of containers these days, for example https://www.slideshare.net/BjrnBackeberg/egi-cloud-container-compute-service}

Makeflow, a well-established workflow management system, has evolved to enable support for application container execution on distributed computing resources~\cite{zheng2015integrating, zheng2017deploying}. In order to capacitate Makeflow with containers, they have conducted a study~\cite{zheng2015integrating} on how to best integrate container technology into workflow systems by analyzing different methods for orchestrating and running workflow tasks (e.g., individual tasks per container, multiple compatible tasks per container, etc.). Our proposed and implemented approach complies with the outcomes of this study, and extends it by providing mechanisms for automated data management for containers.

\section{Conclusion and Future Work}
\label{sec:conclusion}

In this paper, we have described our overall approach to incorporating a variety of container technologies in Pegasus WMS, enabling our users to use them in varied execution environments. Since first releasing support for containers in Pegasus 4.8.0, our user base has slowly and surely started migrating their production workflows to use containers. Our approach focused on efficiently managing automatic container deployment on remote nodes on which workflow jobs are executed. This deployment went hand in hand with ensuring that requisite job input data was made available to the user application in the container while simultaneously preserving data access optimizations (such as mounting directories hosting input data automatically). By working with different user communities, we now realize that the question of whether to execute data transfers required for a job from within the container or from the host OS is not yet answered. We have seen compelling arguments for both approaches. We now feel that, instead of preferring one over the other, we should present the options to the users and Pegasus should support both options regardless.

Thus far, we have allowed users to refer to a pre-built container hosted either in a public container repository or exported as a tar image on a file server. In the future, we plan to allow users to specify the container build file (such as DockerFile in Pegasus catalogs) instead of using a pre-built image. Pegasus will then build the image automatically as part of the executable workflow and deploy the image to remote compute nodes where jobs execute. This makes the user workflow more self contained as the entire environment is described in a Docker file alone, enabling reproducibility and ease of sharing application user workflows. Additionally, we plan to support private container registries by implementing the necessary credential support in our data transfer tool \textit{pegasus-transfer}. We would also like to explore use of Docker alternatives like Podman\cite{podman} which, unlike Docker, is a daemonless container engine used to manage OCI containers and can run containers in root or rootless mode by utilizing Linux namespaces. Use of Podman would enable us to overcome drawbacks of running Docker in HPC environments and would bridge the gap between Docker and Singularity. 

Lastly, the development of Unikernels\cite{2013-unikernels} is of interest, as they are essentially applications compiled along with a library kernel into a micro VM. Packaging in this manner provides  smaller image sizes and faster boot times than containers, and at the same time isolation characteristics of a VM.

% Packaging application code as containers is an attractive proposition,
% Build container as part of workflow
% Enables Reproducibility…. Small mention
% Kubernetes
% Private repositories
% Build container as part of the workflow - makes workflow self-contained with just a Dockerfile/... defining the environment
% We plan to improve docker load in the upcoming Pegasus release
% use section* for acknowledgment
\section*{Acknowledgments}

This work is funded by NSF contract \#1664162, ``SI2-SSI: Pegasus: Automating Compute and Data Intensive Science''; and NSF contract \#1826997, ``CC* Integration: Delivering a Dynamic Network-Centric Platform for Data-Driven Science (DyNamo)''. Development of containerization in PyCBC was supported  by  NSF contract \#1443047, ``CIF21 DIBBs: Domain-Aware Management of Heterogeneous Workflows: Active Data Management for Gravitational-Wave Science Workflows.'' DAB was supported in part by NSF contract \#1748958 to the Kavli Institute for Theoretical Physics.

% trigger a \newpage just before the given reference
% number - used to balance the columns on the last page
% adjust value as needed - may need to be readjusted if
% the document is modified later
%\IEEEtriggeratref{8}
% The "triggered" command can be changed if desired:
%\IEEEtriggercmd{\enlargethispage{-5in}}

% references section

\bibliographystyle{IEEEtran}
\bibliography{bibliography}

% that's all folks
\end{document}